\documentclass[pdflatex,sn-basic]{sn-jnl}

\usepackage{graphicx}%
\usepackage{multirow}%
\usepackage{amsmath,amssymb,amsfonts}%
\usepackage{amsthm}%
\usepackage{mathrsfs}%
\usepackage[title]{appendix}%
\usepackage{xcolor}%
\usepackage{textcomp}%
\usepackage{manyfoot}%
\usepackage{booktabs}%
\usepackage{algorithm}%
\usepackage{algorithmicx}%
\usepackage{algpseudocode}%
\usepackage{listings}%
\usepackage{siunitx}%
\usepackage{multibib} 
\usepackage{hyperref} 
\usepackage{bookmark} 
\newcites{supp}{Supplementary References}  

\begin{document}

\title[Clustering without clusters]{Clustering without clusters: the meta-criterion and centroid reliability mistake continuous dynamics for discrete states}

\author*[1]{\fnm{Frederic} \spfx{von} \sur{Wegner}}\email{f.vonwegner@unsw.edu.au}

\author[2]{\fnm{Gesine} \sur{Hermann}}\email{gesine.hermann@uksh.de}

\affil[1]{\orgdiv{School of Biomedical Sciences}, \orgname{University of New South Wales (UNSW)}, \orgaddress{\street{Wallace Wurth}, \city{Kensington}, \postcode{2052}, \state{NSW}, \country{Australia}}}

\affil[2]{\orgdiv{Department of Neurology}, \orgname{Christian-Albrechts University}, \orgaddress{\street{Arnold-Heller-Strasse 3}, \city{Kiel}, \postcode{24105}, \country{Germany}}}

\abstract{Microstate analysis starts with clustering and the meta-criterion is a heuristic to find an optimum cluster number. We address the following questions: (i) what number is found for non-cluster-forming dynamical systems, (ii) do centroids fall in random attractor regions, (iii) do EEG topographies form clusters or a single connected structure in sensor space?

We find that (i) the meta-criterion suggests spurious optimal cluster numbers (4-8) with high confidence on different attractor geometries, (ii) cluster centroids reliably fall in the same attractor regions, (iii) topological analysis of dynamical systems and resting-state EEG suggests that all form a single connected structure in their respective phase space, not clusters.

We conclude that the meta-criterion should be used with caution and its results should not be taken as representing a ground truth. Cluster numbers deviating from the meta-criterion should not be discarded. 

A more far-reaching implication is that evidence for the existence of clusters in resting-state EEG data is still lacking. Microstate clustering may correspond to the partitioning of a single connected structure.}


\maketitle

\section{Introduction}
\label{sec:intro}

Microstate analysis starts by clustering pre-processed electroencephalography (EEG) data. The usual input to the clustering algorithm are EEG data vectors in sensor space, extracted at local maxima of the global field power time series (GFP peaks) and the cluster centroids are interpreted as characteristic cortical activation patterns \citep{Michel2018}.

While $K=4$ clusters has been the common choice for many years, researchers now explore wider ranges of $K$ which has been stated as an optimization problem, searching for the optimal or 'true' number of clusters \citep{michel2026eeg}. Since K-means and other algorithms used for microstate clustering always return $K$ clusters, even when applied to non-clustered data, other criteria are needed to assess the clustering quality. The meta-criterion was introduced in \citet{Custo2017}, as a 'majority vote' heuristic across 11 different metrics, but has been revised since. The meta-criterion is not a statistical model and therefore provides no p-value or confidence interval. Moreover, the underlying assumption that precedes the clustering step, i.e. the existence of clusters in GFP peak datasets, has not been demonstrated. Preliminary visualization attempts, e.g. two-dimensional PCA projections \citep{vonwegner2026quantitative}, do not readily reveal clusters. 
The two main arguments in favor of the existence of clusters are:
\begin{itemize}
\item \textbf{A1}: Clustering yields highly reproducible cluster centroids instead of random locations in phase space, where the latter might be expected for single structures \citep{michel2026eeg}.
\item \textbf{A2}: A peak in the meta-criterion at $K^*$ is considered evidence for a ground truth of $K^*$ clusters \citep{Custo2017}. 
\end{itemize}

Starting from the hypothesis that GFP peak data might \textit{not} form clusters, and the need for a defined ground truth, we study four dynamical systems that form connected, non-clustered attractors in their respective phase space. We apply the meta-criterion to K-means clustering results from these systems and test the reproducibility of the centroid locations. We then apply topological data analysis (TDA) to the dynamical systems and to real resting-state EEG GFP peak data using a noise-aware clustering method (HDBSCAN) and persistent homology.

We derive conclusions and suggestions for future EEG microstate research.

\section{Materials and Methods}
Details are given in the supplementary material.

\section{Results}
\begin{figure}
\centering
\includegraphics[width=1\textwidth]{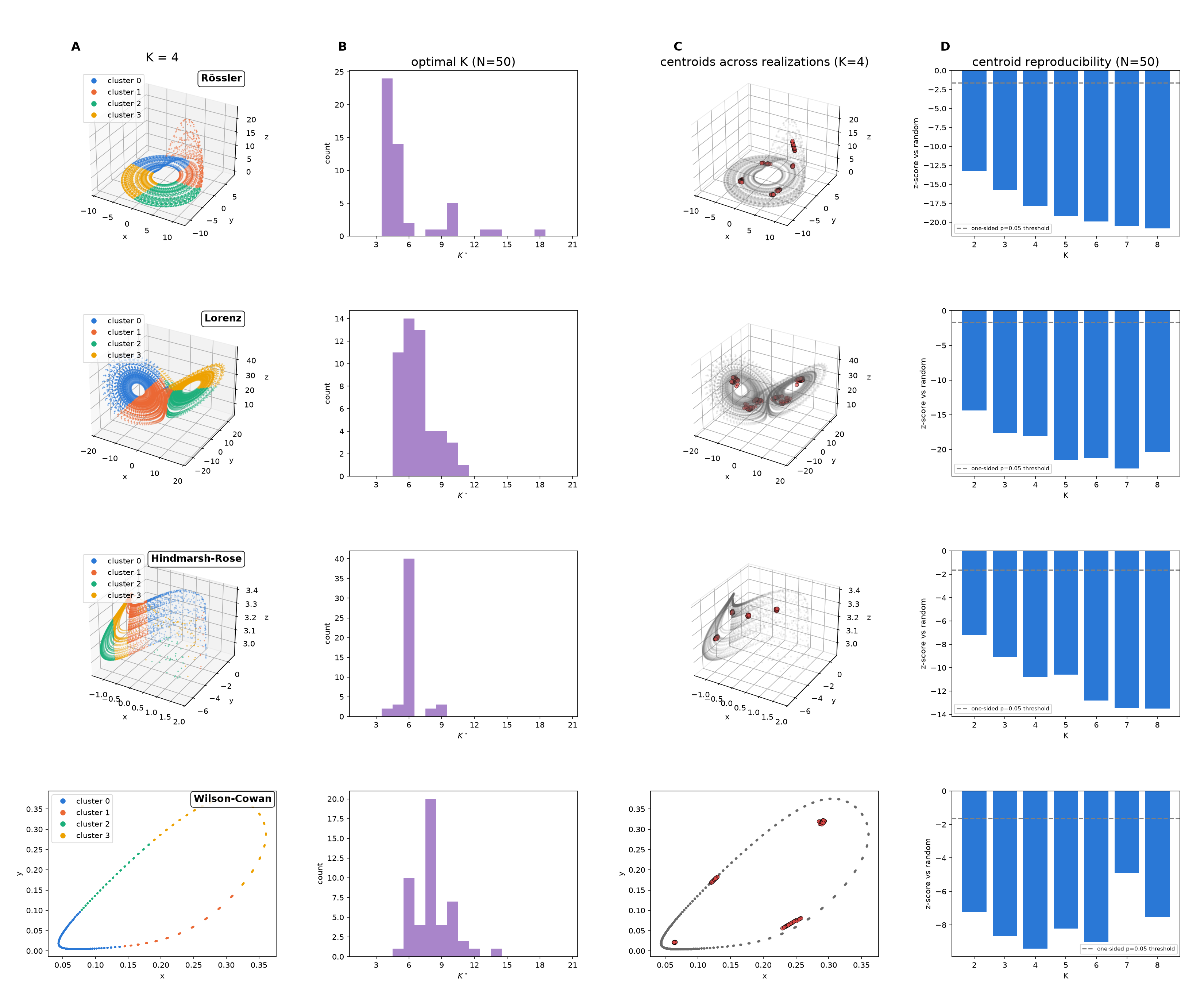}
\caption{Spurious cluster numbers and centroid reproducibility. 
A: Dynamical systems data in their original state space, K-means cluster assignments coloured ($K=4$). 
B: Histogram of optimum cluster numbers from the meta-criterion peak ($N=50$ trials).
C: Localization of cluster centroids ($K=4$) across $N=50$ trials.
D: Centroid reproducibility for $K=2-8$: negative z-scores indicate non-random clustering of centroids on the attractor.
\label{fig:meta-criterion-histograms}}
\end{figure}

Results of applying the meta-criterion to K-means clustering results on dynamical systems are shown in Figure~\ref{fig:meta-criterion-histograms}. Column~A shows each system's attractor, colours representing cluster assignments for $K=4$. Note that K-means draws cluster boundaries through high-density attractor regions. Column B shows histograms of the optimum $K^{\ast}$ suggested by the meta-criterion across $N=50$ randomly initialized simulations. Reproducibility of the cluster centroids is visualized for $K=4$ (C) and analyzed statistically for $K=2-8$ in D, where negative z-scores indicate deviation from randomness. Centroids fall in highly reproducible attractor regions.

\begin{figure}
\centering
\includegraphics[width=1\textwidth]{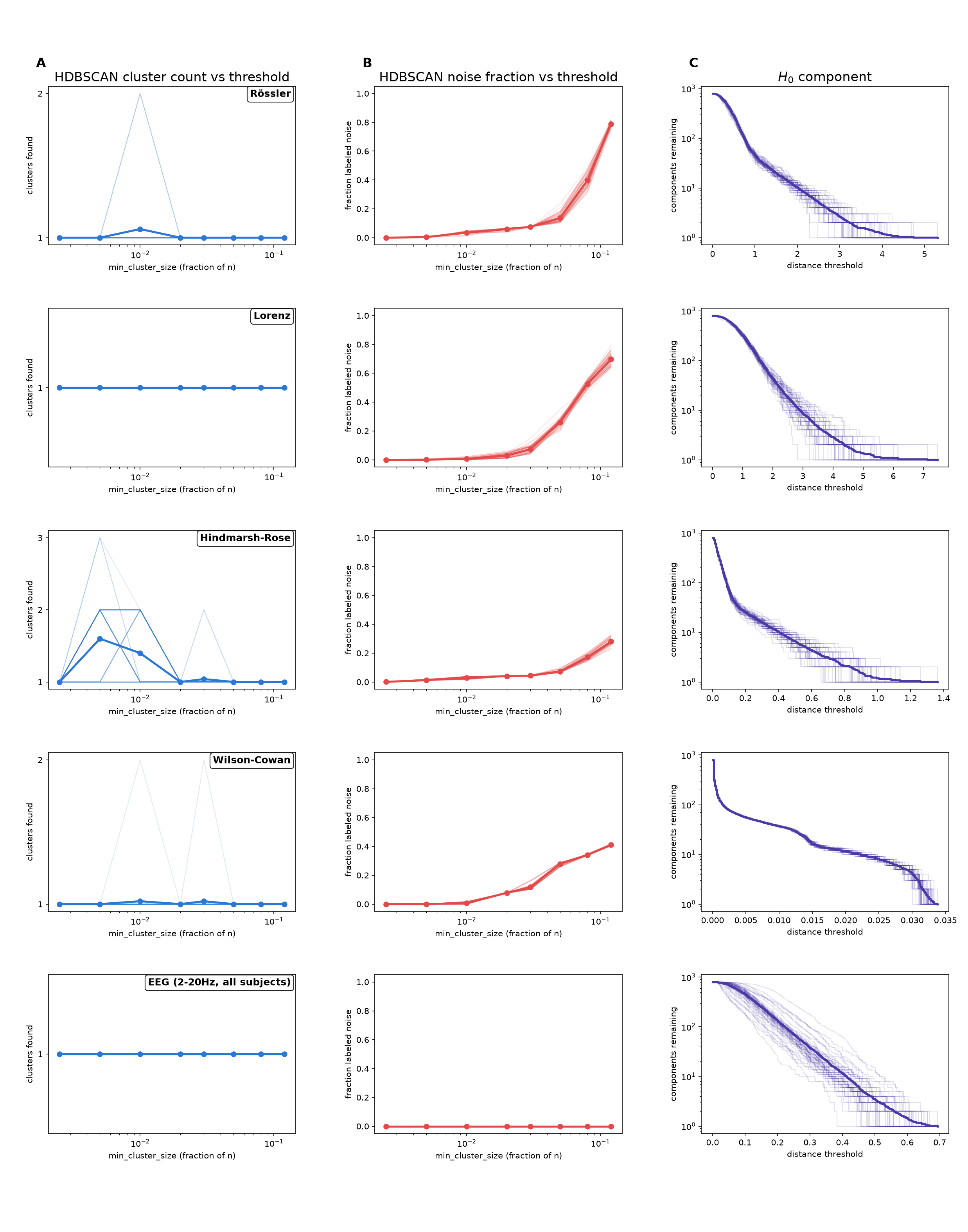}
\caption{Topological analysis of dynamical systems and EEG at GFP peaks.
A, B: HDBSCAN converges to a one cluster solution in all systems, including EEG data (A), assigning some dynamical system data, but not EEG, to noise (B). 
B: Continuously decaying PH component $H_0$ identifies single connected components in all systems, including EEG data.
\label{fig:hdbscan_ph_grid}}
\end{figure}

TDA results are shown in Figure~\ref{fig:hdbscan_ph_grid}. HDBSCAN classification determines a single cluster solution for all systems including a dataset of 61-channel resting-state EEG ($N=125,000$ GFP peaks). Unassigned points increase for dynamical systems but remain at zero for EEG data (B). The $H_0$ component of persistent homology shows a monotone decay for all systems, indicating a single connected structure (C).

\section{Discussion}
\label{sec:discussion}

This study shows that two principal arguments for the existence of clusters in GFP peak data are not self-evident and can be falsified. We provide evidence that (i) a peak in the meta-criterion appears when clustering single connected structures, (ii) cluster centroids fall in highly reproducible attractor regions, far from a random distribution, and (iii) the topological signature of EEG data at GFP peaks is that of a single connected component. Reproducible cluster centroids and meta-criterion peaks should therefore not be regarded as safe indicators of clusters.

We studied systems with known attractor geometry that can be visualized without information loss. Model systems were selected to produce a hierarchy of attractor shapes and degrees of relationship with EEG signals. The meta-criterion identified an unexplained optimum K around 4-8 clusters in all systems. Whether these values reflect an inherent bias of the meta-criterion remains an open question. The meta-criterion has intuitive value as a median-based consensus vote but lacks stronger underpinning in optimization theory. It does not maximize a defined cost function composed of parts addressing the same mathematical concept of clustering and the parts' units are not interconvertible. As the criterion is not based on a statistical model, it does not provide a confidence level or p-value. While using different approaches to what defines a cluster intuitively comes as a strength, making it apparently less likely that they would fail in unison, the numerical results disprove this. This is, to our best knowledge, the first attempt at externally validating this heuristic against a known, non-clustered ground truth.

We provide preliminary evidence for a non-clustered structure of GFP peak data, offering a new perspective on microstates. Direct topological analysis identifies a single connected structure in EEG sensor space, similar to the results for non-cluster-forming dynamical systems. This opens a new line of reasoning about microstates: should GFP peak data be considered a single structure, and is optimal clustering an ill-posed problem?

The results can be reconciled with our current knowledge about neural dynamics. Theoretical and empirical data suggest that neural dynamics evolve as a flow on a lower-dimensional, possibly smooth manifold \citep{Fousek_2024}. If the 'no clusters' hypothesis is true, microstate analysis would still be a valid approach similar to the symbolic dynamics methodology in dynamical systems research \citep{Daw2003}.

\pdfbookmark[1]{Statements and Declarations}{bm:statements}
\section*{Statements and Declarations}
 
\pdfbookmark[2]{Funding}{bm:funding}
\subsection*{Funding}
The authors did not receive support from any organization for the submitted work.
 
\pdfbookmark[2]{Competing Interests}{bm:competing}
\subsection*{Competing Interests}
The authors have no interests to disclose.
 
\pdfbookmark[2]{Ethics approval}{bm:ethics}
\subsection*{Ethics approval}
N/A
 
 
 
 
 
\pdfbookmark[2]{Author contributions}{bm:contributions}
\subsection*{Author contributions}
FvW: software implementation, formal analysis, visualization, writing; GH: concept, evaluation of the meta-criterion on EEG data, writing.

\clearpage

\bibliography{ms_meta-criterion.bib}

\newpage

\pdfbookmark[1]{Supplementary Data}{bm:supp}
\section*{Supplementary Data}
\label{Supp}

\setcounter{figure}{0}
\setcounter{table}{0}
\makeatletter
\renewcommand{\thefigure}{S\arabic{figure}}
\renewcommand{\thetable}{S\arabic{table}}
\renewcommand{\theHfigure}{Supplement.\thefigure}
\renewcommand{\theHtable}{Supplement.\thetable}
\setcounter{section}{1}
\renewcommand{\thesection}{S\arabic{section}}
\renewcommand{\theHsection}{Supplement.\thesection}

\pdfbookmark[2]{Dynamical systems}{bm:dynsys}
\subsection*{Dynamical systems}

\pdfbookmark[3]{Rössler system}{bm:rossler}
\subsubsection*{Rössler system}
The R{\"o}ssler system was introduced as a model system for continuous chaos. We implemented the system as

\begin{align*}
\dot{x} &= -y - z \\
\dot{y} &= x + a y \\
\dot{z} &= b + z(x - c)
\end{align*}

with parameters $a = 0.2$, $b = 0.2$, $c = 5.7$, leading to the standard chaotic spiral regime with a single-lobe attractor \citesupp{Roessler1976}.

\pdfbookmark[3]{Lorenz system}{bm:lorenz}
\subsubsection*{Lorenz system}
The Lorenz system is defined via three coupled non-linear ODEs and was implemented as

\begin{align*}
\dot{x} &= \sigma(y - x) \\
\dot{y} &= x(\rho - z) - y \\
\dot{z} &= xy - \beta z
\end{align*}

with parameters $\sigma = 10.0$, $\rho = 28.0$, $\beta = 8/3$ for the classic chaotic parameter regime producing the two-lobe "butterfly" attractor  \citesupp{Lorenz1963}.

\pdfbookmark[3]{Hindmarsh-Rose single-neuron bursting model}{bm:hr}
\subsubsection*{Hindmarsh-Rose single-neuron bursting model}

The Hindmarsh-Rose system models neuronal action potentials including spiking and bursting dynamics. We implemented the three-dimensional version

\begin{align*}
\dot{x} &= y - a x^3 + b x^2 - z + I \\
\dot{y} &= c - d x^2 - y \\
\dot{z} &= r\big(s(x - x_{\text{rest}}) - z\big)
\end{align*}

where the variables $(x, y, z)$ represent membrane potential, fast current, and slow adaptation current, respectively. We used the parameters $a=1$, $b=3$, $c=1$, $d=5$, $r=0.005$, $s=4$, $x_{\text{rest}}=-1.6$, and stimulation current $I=3.25$ to achieve chaotic bursting. Neurophysiologically, this corresponds to fast-spiking modulated by the slow adaptation variable $z$ \citesupp{Hindmarsh1984}.

\pdfbookmark[3]{Wilson-Cowan excitatory/inhibitory population model}{bm:wc}
\subsubsection*{Wilson-Cowan excitatory/inhibitory population model}
The Wilson-Cowan system has been introduced to model voltage oscillations at the neural mass level, for example, to describe EEG dynamics. It is a two-dimensional system describing the population-level activity of excitatory ($E$) and inhibitory ($I$) pools of neurons. We implemented

\begin{align*}
\tau_E \dot{E} &= -E + S_E(c_1 E - c_2 I + P) \\
\tau_I \dot{I} &= -I + S_I(c_3 E - c_4 I + Q)
\end{align*}

with sigmoid non-linearity baseline-subtracted so $S(0) = 0$: $S(x; a, \theta) = \frac{1}{1 + e^{-a(x - \theta)}} - \frac{1}{1 + e^{a\theta}}$ and the following parameters: $c_1=16$, $c_2=12$, $c_3=15$, $c_4=3$, $a_E=1.3$, $\theta_E=4$, $a_I=2$, $\theta_I=3.7$, $\tau_E=\tau_I=1$, $P=1.25$, $Q=0$. This parameter choice produces limit-cycle dynamics in the $(E, I)$ plane, or voltage oscillations, so all trajectories lie on a smooth closed ring. This is the only non-chaotic model in this study \citesupp{Wilson1972}.

\pdfbookmark[3]{Numerical integration}{bm:numint}
\subsubsection*{Numerical integration}
All four systems were integrated with the Runge-Kutta method from SciPy package (scipy.integrate.solve\_ivp) using the RK45 method and setting relative and absolute tolerance levels to $\mathrm{rtol} = \mathrm{atol} = 10^{-9}$ \citesupp{Virtanen2020}. Initial transients were discarded. Table~\ref{tab:integration-params} lists the numerical integration parameters used for each system. The time interval $\mathrm{dt}$ refers to the sampling interval at which values are recorded, not the internal adaptive step size for integration which determines the solution's accuracy.

\begin{table}[h]
\centering
\begin{tabular}{lllll}
\toprule
System & $dt$ & $t_{\max}$ & discard & default $x_0$ \\
\midrule
Lorenz & 0.01 & 100.0 & 5.0 & $(1,\,1,\,1)$ \\
R\"ossler & 0.1 & 600.0 & 100.0 & $(1,\,1,\,1)$ \\
Wilson-Cowan & 0.05 & 250.0 & 50.0 & $(0.1,\,0.1)$ \\
Hindmarsh-Rose & 1.0 & 5000.0 & 200.0 & $(-1,\,0,\,0)$ \\
\bottomrule
\end{tabular}
\caption{Numerical integration parameters for each dynamical system.}
\label{tab:integration-params}
\end{table}

\pdfbookmark[2]{Electroencephalography (EEG) data}{bm:eeg}
\subsection*{Electroencephalography (EEG) data}
EEG analyses are based on the publicly available LEMON dataset \citesupp{Babayan2019}. We used 61-channel eyes-closed resting-state datasets from 203 subjects. EEG recordings sampled at 250~Hz were band-pass filtered to the 2-20~Hz range. 
For clustering, voltage vectors were sampled at local maxima of the global field power (GFP) time series. In total, we obtained 8,384–13,652 GFP peaks per subject. These were sub-sampled to reduce computational cost. HDBSCAN and the $H_0$ component of persistent homology (PH) were obtained from 50 randomly selected subjects, to approximately match the 50 dynamical systems realizations. Thus, topological data analysis (TDA) results pool statistics from 50 subjects (total cohort size 203). We included 2500 GFP peaks per subject. One exception is Figure~\ref{fig:hdbscan-topomap} which shows the sensor space image of the single HDBSCAN cluster centroid across all 203 subjects (120 peaks/subject × 203 subjects = 24,360 data points). 
Modified K-means clustering of EEG data was conducted in Pycrostates as reported earlier \citesupp{von_Wegner_2025}. 

\pdfbookmark[2]{The meta-criterion}{bm:metacrit}
\subsection*{The meta-criterion}
The meta-criterion was first introduced in a supplementary document to \citesupp{Custo2017supp} and has since undergone several modifications. The original 2017 version contained 11 individual criteria, whereas later implementations used only seven, as described in \citesupp{Brechet_2019} and \citesupp{Brechet_2020}. Our implementation of the \citesupp{Custo2017supp} version misidentified the cluster number on synthetic datasets containing clearly separated Gaussian blobs and consistently produced $K^{*}$ peaks around 10-12 for all tested systems (data not shown). To obtain optimum performance of the meta-criterion, we implemented the most recent version available for this study. This meta-criterion uses seven criteria and is given in the current version of the open-source Cartool repository (\url{https://github.com/DenisBrunet/Cartool}) \citesupp{Brunet_2011}. We implemented the Cartool versions of the individual criteria and the meta-criterion in Python. 

Of note, these are not the same seven criteria as in \citesupp{Brechet_2019,Brechet_2020}. Cartool source code annotations inform us that the Gamma and Dunn criteria used in \citesupp{Brechet_2019,Brechet_2020} were removed, due to systematic overshoot and a lack of contribution to the meta-criterion. Also, Cross-Validation (CV) is currently not enabled as a default. 

The Cartool method to combine the individual criteria is also different from \citesupp{Custo2017supp}. Instead of applying the argmax function to the combined curve, each of the seven curves is ranked independently and its peak identified using argmax. The final value returned by the criterion is the median of those seven peaks. Details are given below.

\pdfbookmark[3]{Shared quantities}{bm:shared-quantities}
\subsubsection*{Shared quantities}
Some quantities are shared by several metrics. In this section, we will use a lower case $k$ to indicate that the variable covers a whole range.
\begin{itemize}
\item K-means is fit once per $k$ over the range $k \in [\max(1, K_{\min}-1),\ K_{\max}+1]$ so that the finite-difference criteria can access the neighboring $k{-}1$ and $k{+}1$ fits.
\item A distance $d(\cdot,\cdot)$ metric between data vectors. The non-polarity-invariant dynamical systems data use the standard Euclidean metric. EEG data vectors are measured with a polarity-invariant metric which is used by the following criteria: Davies-Bouldin, Krzanowski-Lai, Point-Biserial, and Silhouettes. Distances between EEG vectors $u$ and $v$ correspond to the cosine similarity between pre-normalized vectors, which coincides with Pearson correlation for average-referenced EEG topographies. Thus, $d(u,v) = \sqrt{2(1 - \lvert r \rvert)}$, where $r$ is the dot product between $u$ and $v$. Choosing the polarity such that the dot product is positive, i.e. replacing $r$ by $\lvert r \rvert$, is equivalent to $\min(\lVert u - v \rVert, \lVert u + v \rVert)$ for unit vectors. 
\item Three criteria (Point-Biserial, Silhouettes, Cartool version of Krzanowski-Lai) need the full pairwise-distance matrix and are computed on one fixed random subsample of 600 points, drawn once per call and reused across all $k$. Davies-Bouldin is computed on the full dataset directly using point-to-own-centroid distances via the fitted K-means labels and centroids.
\item The $k\ge4$ floor is applied inside each criterion's rank transform $R_c$, and not on a merged curve as in \citesupp{Custo2017supp}.
\end{itemize}

\pdfbookmark[3]{The 7 criteria of the updated meta-criterion}{bm:seven-criteria}
\subsubsection*{The 7 criteria of the updated meta-criterion}

\paragraph{1. Davies-Bouldin}
The Davies-Bouldin criterion is: 

\begin{align*}
\mathrm{DB}(k) &= \frac{1}{k}\sum_{i=1}^k \max_{j \ne i} \frac{S_i + S_j}{M_{ij}}
\end{align*}

where $S_i$ is the median distance of cluster $i$'s points to its centroid and $M_{ij}$ is the distance between centroids $i$ and $j$. This quantity is the only one that requires sign inversion since smaller values indicate better clustering.

\paragraph{2. Davies-Bouldin robust derivative}
This criterion uses the second derivative (discrete elbow) of the rank-transformed Davies-Bouldin values, not of its raw values.

\paragraph{3. Modified Krzanowski-Lai (KL)}
The original KL criterion has been modified in Cartool where it is defined as:

\begin{align*}
\mathrm{sumw}(k) = \sum_{i=1}^{k}\ \sum_{\substack{u,v \in C_i(k) \\ u<v}} d(u,v)^2
\end{align*}

$C_i(k)$ denotes cluster $i$'s points at cluster count $k$ and $d(u,v)$ is the distance metric defined above. Then

\begin{align*}
w(k) = \mathrm{sumw}(k)\cdot k^{2/p}
\end{align*}
where $p$  is the data dimensionality. Setting $d_1(k) = w(k-1) - w(k)$ and $d_2(k) = w(k) - w(k+1)$, the modified KL criterion is

\begin{align*}
\mathrm{KL}(k) = \frac{d_1(k) - d_2(k)}{\max\big(w(k-1), w(k), w(k+1)\big)}.
\end{align*}

\paragraph{4. Point-Biserial}
Let $S$ be the set of data points, $d(u,v)$ the distance metric, and $\ell_k(u)$ data point $u$'s cluster label at cluster number $k$. The Point-Biserial criterion is defined over all pairs $u < v$ in $S$, i.e. count each unordered pair $\left(u,v\right)$ once:

\begin{align*}
\mathrm{PB}(k) = \mathrm{corr}_{\{u,v\}\subset S,\ u<v}\Big(d(u,v),\ \mathbf{1}\big[\ell_k(u)\ne\ell_k(v)\big]\Big)
\end{align*}

i.e. the Pearson correlation, across all $\binom{|S|}{2}$ pairs, between the pairwise-distance vector and a binary indicator that is $1$ for between-cluster pairs and $0$ for within-cluster pairs. A positive correlation means between-cluster pairs are systematically farther apart than within-cluster pairs, indicating good clustering. 

\paragraph{5. Point-Biserial robust derivative}
This criterion is the second-derivative of Point-Biserial's rank-transformed curve.

\paragraph{6. Silhouettes}
Let $S$ be the set of data points, indexed by $i$, and $d(u,v)$ the distance metric. With $C_{\ell_k(i)}$ data point $i$'s own cluster, define

\begin{align*}
a(i) &= \frac{1}{|C_{\ell_k(i)}|-1} \sum_{\substack{j\in C_{\ell_k(i)} \\ j\ne i}} d(i,j) \\
b(i) &= \min_{c \ne \ell_k(i)}\ \frac{1}{|C_c|}\sum_{j\in C_c} d(i,j)
\end{align*}

and

\begin{align*}
s(i) = \frac{b(i)-a(i)}{\max\big(a(i),b(i)\big)}.
\end{align*}

If $i$'s cluster is a singleton, then $s(i)=0$. The Silhouette score is then given by:

\begin{align*}
\mathrm{Sil}(k) = \frac{1}{|S|}\sum_{i\in S} s(i)
\end{align*}

\paragraph{7. Silhouettes derivative}
Second-derivative of the Silhouettes curve, computed on raw values, not on ranks as for the two robust derivatives above.

\pdfbookmark[3]{Rank transform}{bm:rank-transform}
\subsubsection*{Rank transform}
For each criterion, its direction-corrected values across the range of tested $k$ values is rank-transformed, tied values share the same rank (dense ranking). 
Formally, for each criterion $c \in \{ 1,\dots,7 \}$ with direction-corrected raw curve $x_c(k)$, the normalized rank score $R_c(k)$ for $k$ clusters is:

\begin{align*}
R_c(k) &=
\begin{cases}
\dfrac{N_c + 1 - \rho_c(k)}{N_c}, & k \ge 4 \text{ and } x_c(k) \text{ finite} \\
0, & \text{otherwise}
\end{cases}
\end{align*}

where $\rho_c(k)$ is the dense rank of $x_c(k)$ among all finite-valued $x_c(k')$ with $k' \ge 4$. Rank 1 is the largest value, tied values share a rank (no rank number is skipped), and $N_c$ is the number of distinct values in that set. Each criterion's optimum is

\begin{align*}
k_c^{*} = \arg\max_{k} R_c(k).
\end{align*}

\pdfbookmark[3]{Merging into the meta-criterion}{bm:merging}
\subsubsection*{Merging into the meta-criterion}
The final answer of the meta-criterion as defined in Cartool combines the individual criteria $k_c^{*} = \arg\max_{k} R_c(k)$ as:

\begin{align*}
K^{*} = \mathrm{round}\big(\mathrm{median}\{k_1^{*},\dots,k_7^{*}\}\big)
\end{align*}

\pdfbookmark[2]{Meta-criterion on a single run of the dynamical systems}{bm:metacrit-example}
\subsection*{Meta-criterion on a single run of the dynamical systems}
\begin{figure}[h]
\centering
\includegraphics[width=1\textwidth]{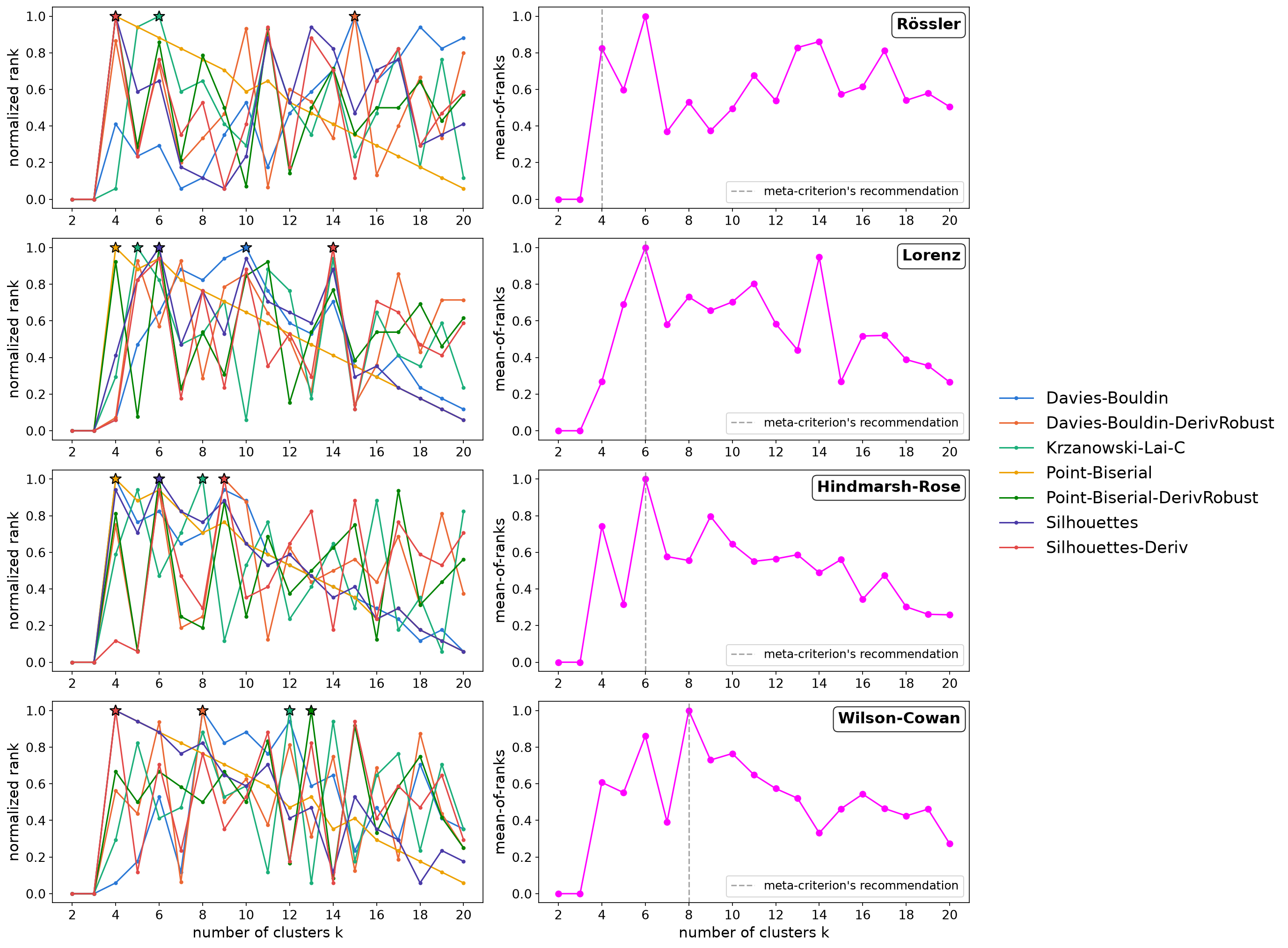}
\caption{The meta-criterion applied to a single realization of each of the four dynamical systems. 
Left: Individual criteria, their top rank indicated by stars.  
Right: Meta-criterion with optimal $K^{\ast}$ indicated by the dashed vertical lines. The curve is the geometric mean of the seven rank scores (one per criterion).
\label{fig:meta-criterion-examples}}
\end{figure}

Figure~\ref{fig:meta-criterion-examples} shows an example of the meta-criterion applied to a single realization of each of the four dynamical systems. Individual criteria are shown on the left, with their top rank highlighted (star), and the meta-criterion is on the right, its final recommendation shown by the dashed vertical line. The curves on the right are for display only, they do not enter the calculation of $K^{*}$, and represent the geometric mean of all seven rank scores.

\pdfbookmark[2]{Meta-criterion on EEG data}{bm:metacrit-eeg}
\subsection*{Meta-criterion on EEG data}
\begin{figure}[h]
\centering
\includegraphics[width=1\textwidth]{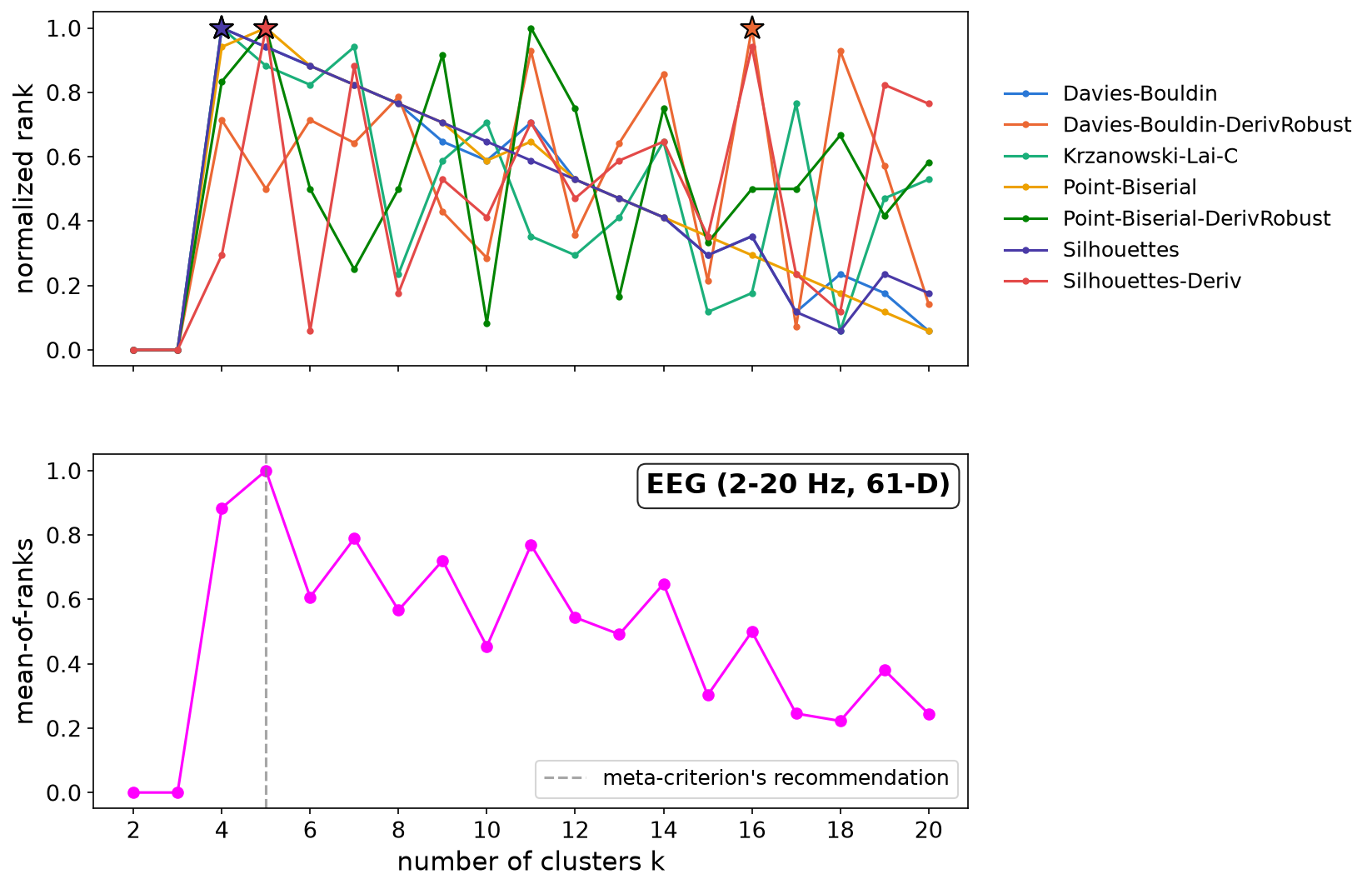}
\caption{The meta-criterion applied to the EEG dataset. Top: Individual criteria and top ranks (stars). Bottom: geometric mean of rank scores (curve) and final recommendation of $K^{\ast}=5$ (vertical dashed line).
\label{fig:meta-criterion-eeg}}
\end{figure}

Although this study is not a recommendation to apply the meta-criterion to microstate clustering, we show the results on our EEG dataset in Figure~\ref{fig:meta-criterion-eeg} for completeness. The meta-criterion's recommendation was $K^{\ast}=5$ in this case.

\pdfbookmark[2]{Centroid spatial reproducibility}{bm:centroid-repro}
\subsection*{Centroid spatial reproducibility}
To test whether K-means centroids fall in non-random locations on the attractor, we compared the observed spatial distribution of K-means centroids across independent model realizations to a null model of randomly chosen attractor points. For each of $N$ independently simulated trajectories drawn from random initial conditions, K-means was fit once, yielding $N$ sets of $K$ centroids. Centroids were pooled across realizations, and the test statistic was the mean cross-run nearest-neighbor distance: for each centroid, the Euclidean distance to its nearest neighbor among centroids from a different realization. The null distribution was built by drawing, for each of 2,000 resamples, $K$ points uniformly at random from each of the same $N$ realized trajectories and computing the identical cross-run statistic on these surrogate points.
Computing the null from actual attractor points is essential since the attractors occupy small, non-uniform, and lower-dimensional volumes of the phase space. 
Significance was assessed as a one-sided empirical $p$-value and summarized as a $z$-score relative to the null mean and standard deviation.

\pdfbookmark[2]{Topological data analysis (TDA)}{bm:tda}
\subsection*{Topological data analysis (TDA)}

\pdfbookmark[3]{HDBSCAN clustering}{bm:hdbscan}
\subsubsection*{HDBSCAN clustering}
Hierarchical Density-Based Spatial Clustering (HDBSCAN) was chosen as a comparison to K-means as it has two advantages over other clustering algorithms \citesupp{McInnes2017}. First, the algorithm can converge to a single ($K=1$) and even a zero ($K=0$) cluster solution. Second, it can leave data points unassigned, these are usually labelled as 'noise'. Noise is not to be interpreted in the context of the original data set. In our case, it does not mean that the corresponding segment of the EEG recording is noisy in an electrophysiological sense.  

HDBSCAN was run across a range of minimum cluster size thresholds (0.25-12\% of $n$), recording the number of clusters found (excluding noise) and the fraction of points labeled noise at each threshold. The full scan was repeated across $N$ independent realizations, and both curves (cluster count and noise fraction) were pooled to assess whether a multi-cluster structure or a unimodal continuum with a shrinking core would emerge. The threshold $\epsilon$ for cluster selection was set to 0.35 times each dataset's median pairwise distance. This prevents HDBSCAN from partitioning near-one-dimensional manifolds into micro-clusters at fine thresholds.

\pdfbookmark[3]{Persistent homology}{bm:ph}
\subsubsection*{Persistent homology}
Persistent homology (PH) is not a clustering algorithm but an instrument for TDA rooted in algebraic topology, as implemented in the ripser Python software package \citesupp{Tralie2018}. For our purpose, we only need the $H_0$ part that counts the number of connected components in spaces of arbitrary dimension.
$H_0$ was computed across different scales using the 'barcode' approach of the Vietoris-Rips filtration, computed on random subsamples (n=800) of each dataset. $H_0$ persistence is mathematically equivalent to single-linkage hierarchical clustering, with each finite death time marking the distance threshold at which two components merge. Formally, $H_0$ is a group and the reported "$H_0$ component" is the group's rank, also known as the 0th Betti number $\beta_0(\varepsilon)$, as a function of the distance threshold $\varepsilon$. We tracked the full component-count curve starting near $n$ and stepping down to one as thresholds increase, with each step corresponding to one merge. 

Genuine, well-separated cluster structures produce a plateau as shown in Figure~\ref{fig:hdb_ph_blobs}, a range of thresholds over which the component count remains constant at $K$, whereas a continuum decays smoothly with no sustained plateau, as seen in Figure~\ref{fig:hdbscan_ph_grid}~C. As with HDBSCAN, curves were pooled across $N$ independent realizations. Samples with fewer points than the subsample target were evaluated via step-function lookup on a shared distance grid rather than interpolation, keeping curves of different underlying size directly comparable. 

\pdfbookmark[2]{Metrics}{bm:metrics}
\subsection*{Metrics}
For EEG data, HDBSCAN and PH used the polarity-ignoring squared-cosine distance 

\begin{align*}
d(u,v) &= 1-\left(\dfrac{u \cdot v}{\lVert u \rVert \lVert v \rVert}\right)^2
\end{align*}

in line with the common understanding that topographies with inverse polarity are considered identical. The same principle is used in the modified K-means algorithm \citesupp{Pascual-Marqui1995}.

\pdfbookmark[2]{Additional controls}{bm:controls}
\subsection*{Additional controls}
\begin{figure}[h]
\centering
\includegraphics[width=0.5\textwidth]{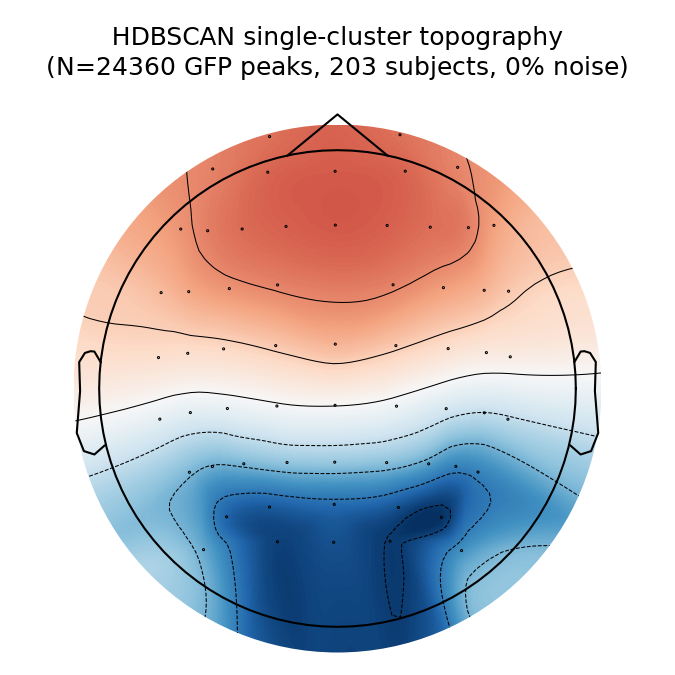}
\caption{Sensor-space topography of HDBSCAN's single cluster, pooled across all 203 EEG subjects (24,360 GFP peaks, 0\% noise).
\label{fig:hdbscan-topomap}}
\end{figure}

We performed two additional control experiments ('sanity checks'). 
First, we projected the centroid of the single cluster detected by HDBSCAN from GFP peaks pooled across all subjects back into EEG sensor space. The result is shown in Figure~\ref{fig:hdbscan-topomap} and can be interpreted either as microstate C or a typical alpha band power distribution. Thus, the identified cluster is meaningful in a neurophysiological sense and cannot be considered an artefact.

\begin{figure}[h]
\centering
\includegraphics[width=0.8\textwidth]{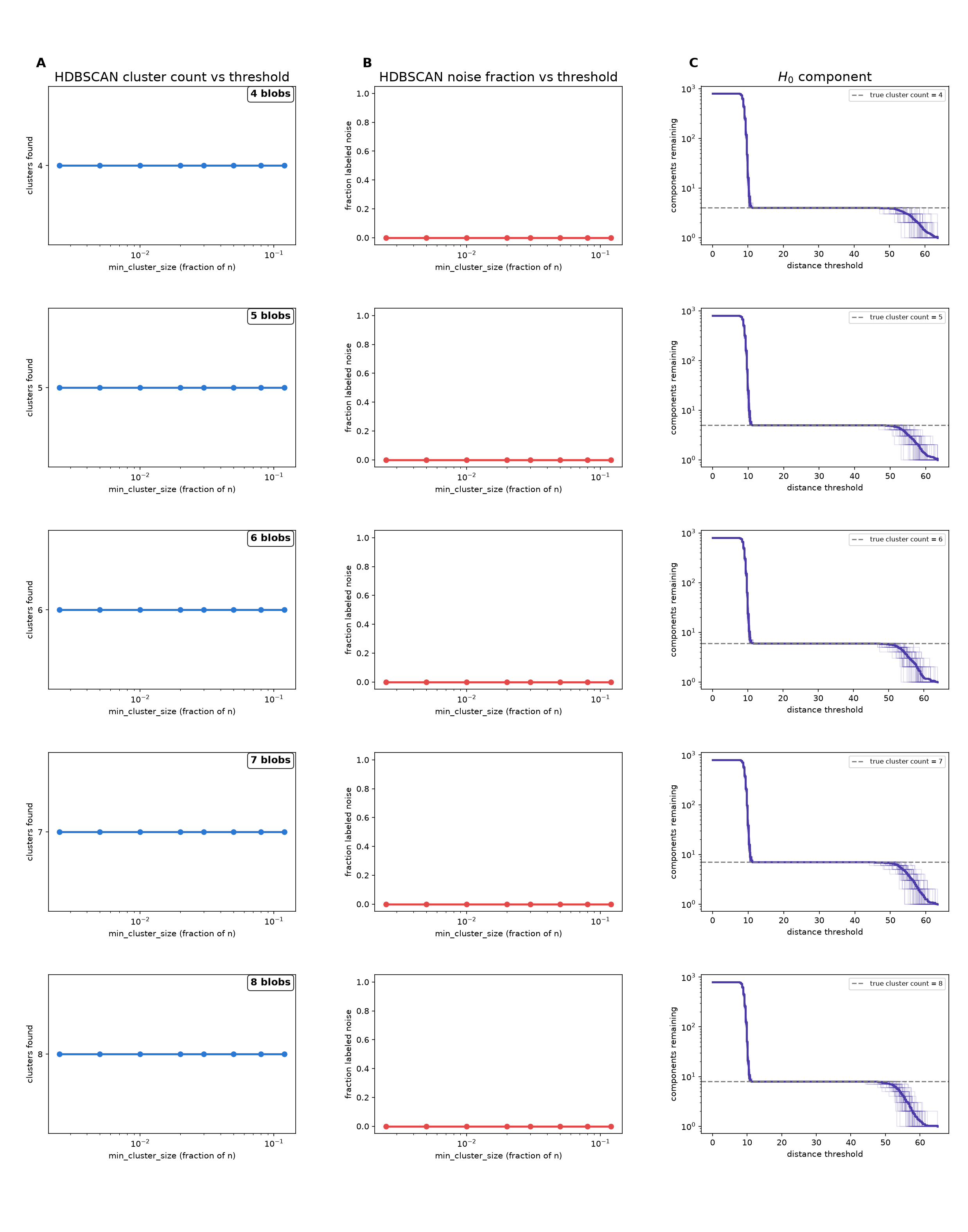}
\caption{HDBSCAN and PH results for synthetic cluster data with a known ground truth of $K=4-8$ 61-dimensional Gaussian blobs. HDBSCAN recovers the correct cluster number across a wide range of cluster size thresholds (A) and leaves no data points unassigned (B). The $H_0$ component of PH produces clear plateaus at the correct cluster number.
\label{fig:hdb_ph_blobs}}
\end{figure}

Second, to illustrate that HDBSCAN and PH are capable of detecting real clusters, Figure~\ref{fig:hdb_ph_blobs} illustrates the results for $K=4-8$ synthetic Gaussian blobs in 61 dimensions, matching our EEG data dimensionality. HDBSCAN correctly identified clusters and left no points unassigned. PH curves show a clear plateau at the level indicating the correct cluster number (dashed line) across a wide range of distance thresholds.

\clearpage
\begingroup
\let\refname\refnamesupp
\let\bibname\refnamesupp

\endgroup

\end{document}